# One-Pot Approach for Acoustic Directed Assembly of Metallic and Composite Microstructures by Metal Ion Reduction

*Avraham Kenigsberg, Heli Peleg-Levy, Haim Sazan, Silvia Piperno, Liron Kenigsberg, Silvia Piperno & Hagay Shpaisman\**

Department of Chemistry and Institute of Nanotechnology and Advanced Materials, Bar-Ilan University, Ramat Gan 5290002, Israel

E-mail: hagay.shpaisman@biu.ac.il

ABSTRACT

Acoustic-directed assembly is a modular and flexible bottom-up technique with the potential to pattern a wide range of materials. Standing acoustic waves have been previously employed for patterning preformed metal particles, however, direct patterning of metallic structures from precursors remains unexplored. Here, we investigate utilization of standing waves to exert control over chemical reaction products, while also exploring their potential in the formation of multi-layered and composite micro-structures. Periodic metallic micro-structures were formed in a single step, simplifying microstructure fabrication. Concentric structures were obtained by introducing a metal precursor salt and a reducing agent into a cylindrical piezoelectric resonator that also served as a reservoir. In addition, we introduce an innovative approach to directly fabricate metallic multi-layer and composite structures by reducing different metal ions or adding nanoparticles during the reduction step. Fewer steps are needed, compared with other methods, and there is no need to stabilize the nanoparticles or to ensure chemical affinity between the metallic matrix and inorganic nanoparticles. This innovative approach is promising for production of complex microstructures with enhanced functionality and controlled properties.

INTRODUCTION

Micro-patterning of metals is highly desirable for applications in electronics,[1] sensing,[2] catalysis, and optics.[3,4] Mesoporous micro-structures are of particular interest when a high surface area is required.[5,6] Metal-based composite materials with enhanced physical or chemical properties are used to improve mechanical,[7] electrical [8] or magnetic properties,[9] for hydrogen storage, [10] catalysis [11,12] and sensing applications.[13,14] Microstructures can be fabricated by top-down or bottom-up approaches. While top-down methods such as photolithography offer high resolution in micro and nano scales, bottom-up assembly methods allow complex structures with significantly fewer steps and material waste.

Bottom-up techniques include inkjet printing,[15] laser induced forward transfer (LIFT),[16] directed laser writing,[17] and magnetic or electric field directed assembly.[18] Each method has its own advantages and limitations. Some require preformed particles or are limited to a certain feature size. Others require specific physical properties and are limited to a certain material group. A versatile bottom-up fabrication technique that is suitable for multiple types of materials presents a significant challenge.

Acoustic directed assembly is a versatile technique that can be used to fabricate various types of materials.[19–22] Its main advantages are modularity and flexibility due to minimal requirements from particles and host fluid. This technique works simultaneously on the required pattern rather than sequentially as other printing methods. Suspended particles subjected to an acoustic radiation force associated with a standing wave migrate to specific locations [23–25] (nodes or antinodes) determined by compressibility and density difference between the particles and host fluid. Particles from nano to supra-millimetric scales can be trapped and manipulated.[26,27] Prior studies have primarily focused on biological cells, assembling particles in continuous flow systems, and demonstrating holographic techniques by assembling polymeric particles.[28–34]

Standing acoustic waves were shown to pattern preformed metal particles, [19,35] colloidal particles[20] and cells.[22,36,37] However, acoustic directed patterning of metallic structures directly from metal ions was not realized. Here, such assembly of metallic microstructures is shown by chemical reduction and acoustic direction in a single step. Composite structures are fabricated by combining two types of ions or adding nano-additives to the growing structure in the presence of acoustic waves.

By reducing metal ions while standing bulk acoustic waves (SBAWs) are present, zero valent NPs are simultaneously formed and directed to specific locations by the acoustic field, resulting in continuous micro-patterns. This one-pot approach does not require stabilization or special treatment to allow aggregation of NPs. We studied the effect of acoustic waves with uneven force distribution generated in a cylindrical reservoir[38] on the geometry of the micro-pattern. The effect of various source amplitudes on ring geometry and overall deposition appearance was also studied. We characterized the ring continuity by conductivity measurements and examined the effect of the wave on the meso-structure by cross-section analysis. The influence of acoustic force on microstructure formation of single layers, multiple layers, and hybrid microstructures was investigated. This method has the potential to become a new tool for fabrication of multilayered microstructures containing various types of materials.

EXPERIMENTAL SECTION

Materials

Silver nitrate and sodium hydroxide were obtained from Fischer Scientific, sodium borohydride from Acros, gold (III) chloride hydrate from Alfa Aesar, iron (III) oxide nano powder (particle size <50 nm) from Sigma Aldrich, and isopropyl alcohol from Carlo Erba.

Acoustic directed assembly setup

Lead zirconate titanate (type-II, Americanpiezo) piezoelectric cylinders were used with inner diameter of 22 mm, outer diameter of 26 mm and height of 20 mm. As the cylinders also serve as reservoirs, they were attached to glass microscope slides by silicone-based high vacuum grease (Dow Corning) to prevent liquid leakage (Figure 1). For cross-sectioning, single-side polished silicon wafers (Universitywafer, USA) were cut into ~3×4 cm. These assemblies were placed on the stage of an inverted microscope (Nikon Eclipse TS100). To protect the inner side of the resonator from aggressive chemicals, a protective polydimethylsiloxane (PDMS) layer was deposited by dipping the PZT resonator in Sylgard TM184 silicon elastomer base and curing agent (Dow Corning) mixed at a 10:2 weight ratio. The PDMS coating was dried at room temperature for 20 min, leaving a thin (70–100 µm) uniform well-adhered layer (most of the coating was removed due to gravitation), and cured at 80°C for 3 h.

The resonator was activated around its resonance frequency (1.1 MHz) by a function generator (Siglent, SDG 5162) at 10 Vpp (unless stated otherwise), creating pressure waves inside the reservoir. A power amplifier (X2, FYA2030, 3 MHz bandwidth with low distortion) was used to enhance the signal to 20 Vpp at 50 Ω unless stated otherwise. To determine the optimal resonance frequency of each resonator, 2 ml of 0.02 wt% polystyrene (2 µm) in deionized (DI) water dispersion was added and frequencies were swept between 1 and 1.2 MHz at 1 kHz steps. The shape and uniformity of the polystyrene concentric ring patterns were evaluated using an inverted optical microscope. The optimal frequency was defined as the frequency that formed the most well-defined and continuous concentric patterns.

Fabrication of micro-patterns

Silver and gold micro-patterns were fabricated in piezoelectric cylinders filled with a total of 2 ml DI water-based solutions containing silver nitrate (4.4 mM) or gold (III) chloride hydrate (2 mM) followed by addition of sodium borohydride and sodium hydroxide (1 and 1.4mM for silver, 3 and 20mM for gold, respectively). Acoustic waves were applied after silver nitrate/ gold chloride addition (prior to borohydride and hydroxide addition).

Fabrication of layered silver/gold micro-structures process was initiated by creating a silver micro-pattern. The piezoelectric cylinder was filled with a 2 ml solution of deionized water containing silver nitrate (4.4 mM), sodium borohydride (1 mM), and sodium hydroxide (1.4 mM). Acoustic waves were applied following the addition of silver nitrate prior to borohydride and hydroxide addition. After 30 minutes, the wave was deactivated, and the aqueous medium was removed. Next, a gold micro-pattern was generated atop the silver layer. The same piezoelectric cylinder was filled once more with a 2 ml solution of deionized water comprising gold chloride (2 mM), sodium borohydride (3 mM), and sodium hydroxide (20 mM). Again, acoustic waves were applied after the addition of gold chloride.

Silver + gold bimetallic micro-patterns were fabricated by adding the silver and gold solutions to the same reservoir and applying SBAWs for simultaneous reduction with sodium hydroxide and borohydride. The piezoelectric cylinder was filled with a total of 2 ml solution of deionized water containing silver nitrate (4.4 mM) and gold chloride (2 mM). Acoustic waves were applied followed by addition of sodium borohydride (4 mM), and sodium hydroxide (21.4 mM). Hematite combined with either silver or gold forming a composite microstructure were fabricated by introducing a suspension of iron (III) oxide nanoparticles (NPs) at a concentration of 0.2 wt% in water into the previously mentioned solutions of gold chloride or silver nitrate ions. The piezoelectric cylinder was filled with a 2 ml solution of deionized water containing silver nitrate (4.4 mM) with 0.2 wt% iron (III) oxide NPs or gold chloride (2 mM) with 0.2 wt% iron (III) oxide NPs. Acoustic waves were applied followed by addition of sodium borohydride and sodium hydroxide at concentrations of 1 mM and 1.4 mM, respectively, for the silver and iron NPs combination, and 3 mM and 20 mM, respectively, for the gold and iron NPs combination.

Multilayered silver/hematite/silver micro-patterns were formed by reduction of a silver layer as described above followed by iron oxide suspension arrangement and reduction of another silver layer on top of the silver/hematite micro-pattern. The piezoelectric cylinder was filled with a 2 ml solution of deionized water containing silver nitrate (4.4 mM), sodium borohydride (1 mM), and sodium hydroxide (1.4 mM). Acoustic waves were applied following the addition of silver nitrate. After 30 minutes, the wave was deactivated, and the aqueous medium was removed. 2ml of 0.2 wt% iron oxide suspension were added to the acoustic reservoir and the wave was activated. After an additional 30 minutes the wave was deactivated, and the aqueous medium was removed again, and an additional silver layer was applied as described before.

Characterization

Bright-field microscopy images were obtained by a Nikon Eclipse TS100 inverted microscope equipped with Nikon objective lenses: 4× (NA 0.13) and 10× (NA 0.30). A full high-definition microscope camera (HDMI16MDPX, DeltaPix) was used to acquire images. SEM images were obtained using a Quanta FEG 250 system (HV 5-15KV) equipped with energy-dispersive X-ray spectroscopy (EDS). Pattern dimensions were measured by a profilometer (Veeco Dektak 150 system; stylus radius 12.5 µm, force 0.5 mg; resolution 0.074 µm/sample). Cross-sections of microstructures were formed by a focused ion beam (FIB) system (Helios 600) or a cleaver (LatticeAx 120). Resistance measurements were performed at a SUSS MicroTec probe station.

RESULTS AND DISCUSSION

Cylindrical piezoelectric resonators (1.1±0.1 MHz) that also serve as reservoirs for the ion solutions were used to produce acoustic waves. Each resonator was attached to a glass cover slip using silicon grease (**Figure 1** a) to prevent leakage. Resonators were filled with an aqueous solution of metal ions and driven by a function generator (1–10 Vpp at 50 Ω) to create SBAWs. Following activation of the waves, the reducing agent (sodium borohydride) was added. Concentric ring-shaped structures were observed almost immediately (Figure 1 b) due to reduction to zero valent metal NPs driven toward nodal areas. Single layers, multilayers, and hybrid structures were formed (summarized in Table 1).

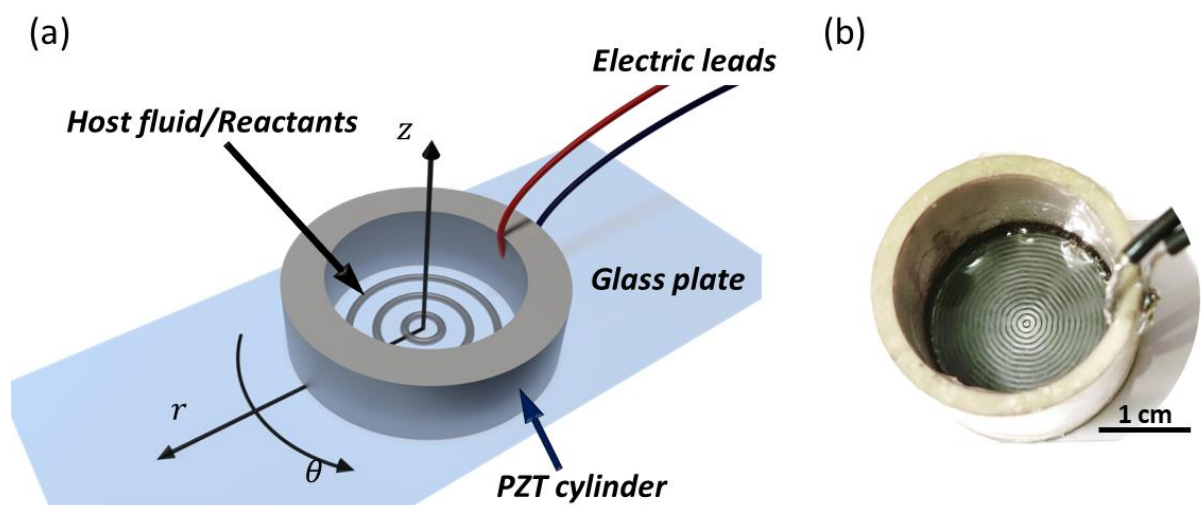

**Figure 1.** (a) Illustration of the experimental apparatus and (b) concentric pattern of zero valent Ag fabricated by acoustic manipulation of silver ion reduction products.

**Table 1.** Acoustic directed microstructures.

| Single layer | Multi-layer | Composite layer |
| --- | --- | --- |
| Gold | gold/silver | gold + silver |
| silver | hematite/silver | silver + hematite |
|  | silver/hematite/silver | gold + hematite |

Single layer

The suggested mechanism for concentric pattern formation is illustrated in **Figure 2**. The reservoir is filled with an aqueous solution of metal ions ($HAuCl_4$ or $AgNO_3$). Sodium borohydride and sodium hydroxide are added immediately after activation of the acoustic wave. Reduction to zero valent metals and formation of NPs by nucleation and growth mechanism are rather straightforward and based on well-known chemical reactions.[39,40]

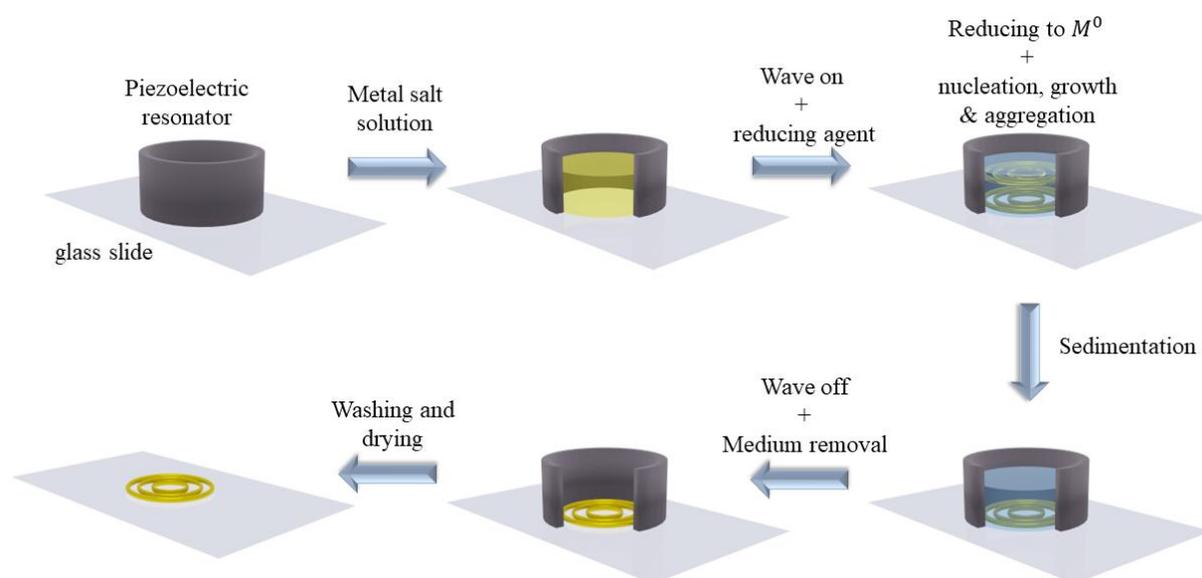

**Figure 2.** Illustration of concentric ring-shaped metal micropattern fabrication steps of a single metal guided by standing acoustic waves.

NPs are driven by the acoustic field toward areas with no acoustic force (pressure nodes). Concentric patterns are visible within less than 10 s. The uncoated NPs aggregate spontaneously and sediment gradually within a few minutes on the glass substrate as concentric rings (video SV1). These patterns are easily observed even without magnification (Figure 1 b). Less than 3 min after adding the reducing agents, no detectable changes were observed (even with the aid of a microscope). To ensure complete fixation, solutions were exposed to SBAWs for another 30 min. The aqueous medium was gently removed by a syringe and dried on a hot plate at 70°C for 10 min. After drying, the acoustic resonator was detached from the glass slide, and remaining residues were washed with water and isopropyl alcohol followed by additional drying (1 min on the hot plate).

The acoustic radiation force for particles with a positive contrast factor in cylindrical acoustic resonators was presented by Barmatz et al..[24,25] Following this expression (Equation S1), the normalized acoustic radiation force for a system containing gold spherical particles dispersed in water as a host fluid was calculated (up to a radius of 7 mm) and plotted as shown in **Figure 3**c. The direction of the radiation force is indicated by purple vectors, and the first nodal area is marked with a yellow dot. Light gray concentric lines are an experimental image of Au deposits; the picture was tilted to demonstrate the good fit between calculated locations of nodal areas and metal deposits. Moreover, nodal areas that have stronger acoustic fields directing to their locations are expected to acoustically trap more material. Indeed, our observations support this hypothesis as shown in Figure 3. The highest peaks of Ag and Au concentric patterns are located close to the central axis, which corresponds to the strongest acoustic forces directing to the first nodal area. From there towards the reservoir edge, a constant decrease in peak height is observed (Figure 3a, d). The decrease in acoustic strength also leads to weaker ordering of formed NPs towards the reservoir edge (see **Figure S1**). The effect of acoustic radiation force on the height of a single ring was studied for Ag and Au depositions (Figure 3b) by changing the input amplitude. Images of concentric patterns fabricated at different amplitudes demonstrating the impact of applied acoustic force on deposition outcome are displayed in Figure 3e and f. By employing different harmonic frequencies (fundamental, second, and third) on the same resonator, we can adjust the quantity of rings along the reservoir's radii (**Figure 4** f). As the frequency increases, both the rings' thickness and the gaps between them become narrower as expected. Furthermore, when the reducing agents are abundant, the width of the patterns is controlled by the concentration of gold ions (**Figure S5**). Furthermore, alternative setups that generate SBAWs can be employed. For instance, refer to **Figure S6**, which demonstrates straight lines rather than rings. Highly intricate designs can be attained either through acoustic holographic patterning[29,31,41] or by utilizing an innovative acoustic stencil technique. [42]

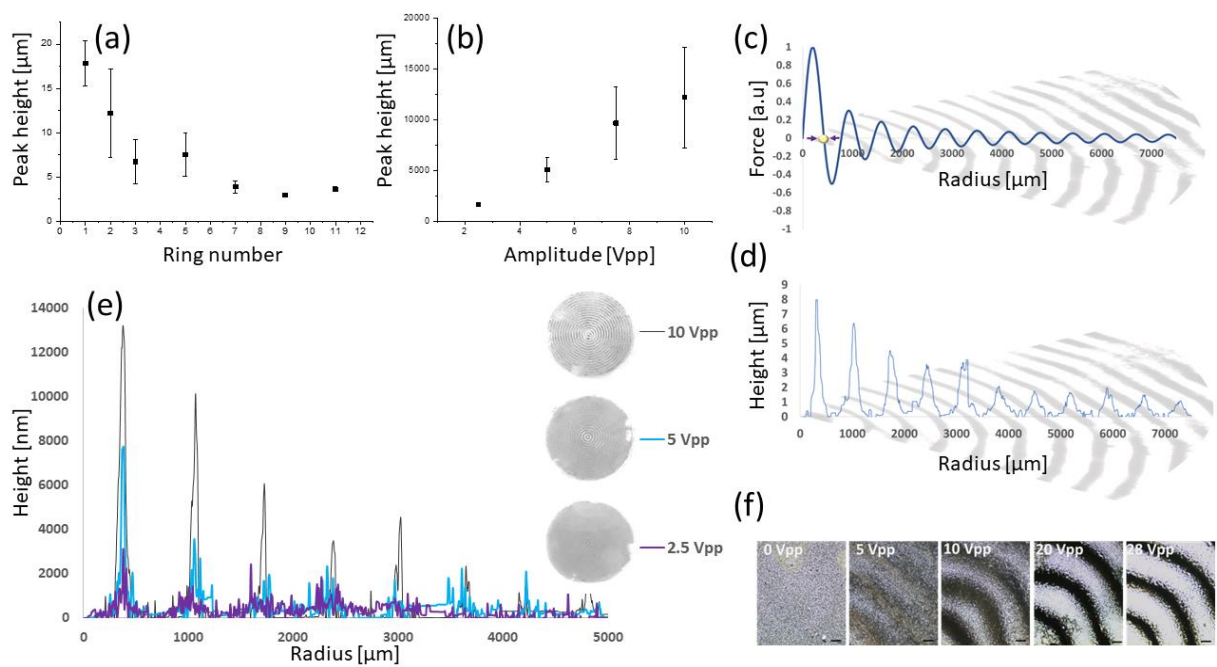

**Figure 3.** (a) Average peak heights of gold ring patterns as a function of ring number (from center). (b) Average peak height of the second gold ring as a function of acoustic wave amplitude. (c) Calculated radiation force inside the PZT resonator for a positive acoustic contrast factor (r = 0 corresponds to the reservoir center). The yellow dot represents a nodal area, and purple arrows indicate force direction. Light gray patterns are experimental microscope images of Au deposits where the picture is tilted to highlight correlation with calculations. (d) Experimental profile measurement of formed metallic rings as a function of radius. (e) Profile measurements of concentric silver patterns formed with 2.5 (purple), 5 (blue) and 10 (black) VPP. (f) Concentric silver micropatterns fabricated in the presence of SBAWs at voltage amplitudes of 0 (reference), 5, 10, 20, and 28 Vpp. Scalebars= 200μm.

SEM and EDS measurements were performed on patterns from gold ion reduction. Figure 4a shows a silver ring deposited on a glass cover slip. EDS measurements (Figure 4b–e) of the patterned area confirm that the pattern consists of Ag, while the silicon and oxygen signals in the background are attributed to the glass substrate.

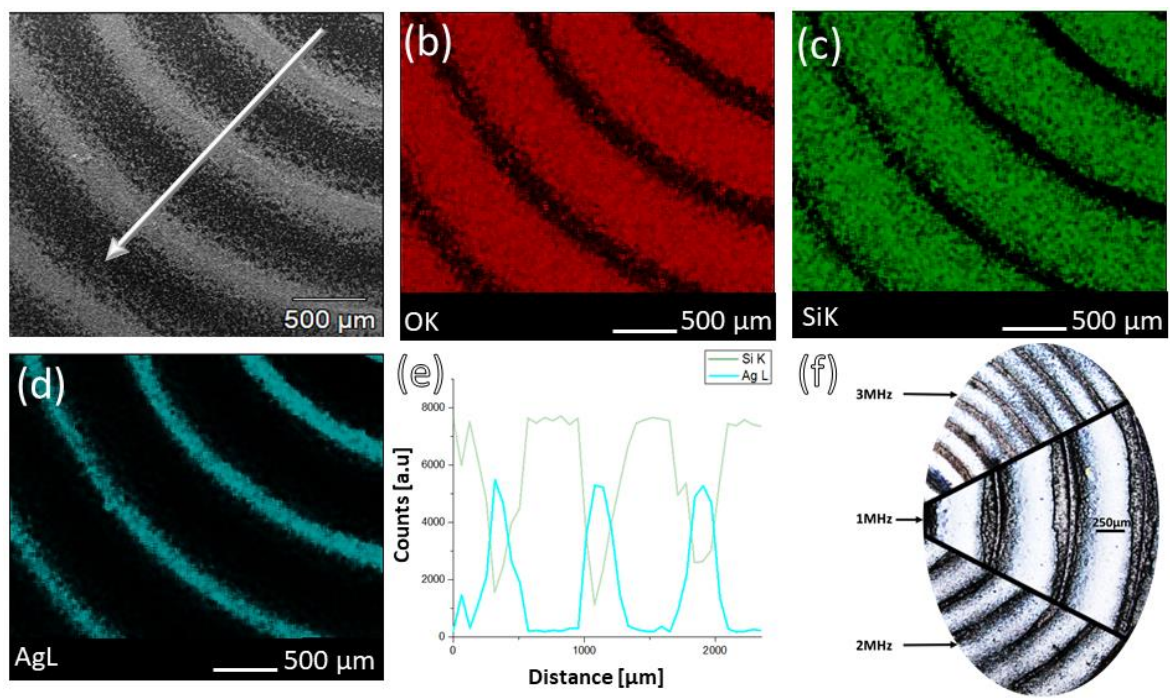

**Figure 4.** (a) SEM image displaying silver rings on a glass substrate, along with EDS mappings of (b) Oxygen, (c) Silicon, and (d) Silver, as well as (e) quantities of Silver and Silicon along the gray arrow. (f) Bright field microscopy (patched) images of silver microstructures fabricated by the same resonator working at its first, second and third resonance frequency.

HR-SEM images were acquired to reveal the mesoporous structure of formed micro-patterns. The samples shown in **Figure 5** were fabricated by reducing Au ions with NaBH$_4$ in alkaline solution. A focused ion beam (FIB, Ga ions as source) was used to expose cross-sections of Au deposits formed with or without acoustic waves (Figure 5b and d). This comparison demonstrates the effect of SBAWs on structure density. A substantial decrease in pore size was observed in the sample fabricated with SBAWs compared to the reference sample. The density of the reference sample has large variations (as can be observed in Figure 5c), Although we chose to analyse areas that appear to have less porous variations from a top view, the mesostructure of the gold is still considerably less dense and with larger pore size variations than the sample fabricated with SBAWs (Figure 5b). The reference sample contains twice the amount of all ingredients to compensate for the larger deposition area without ring formation.

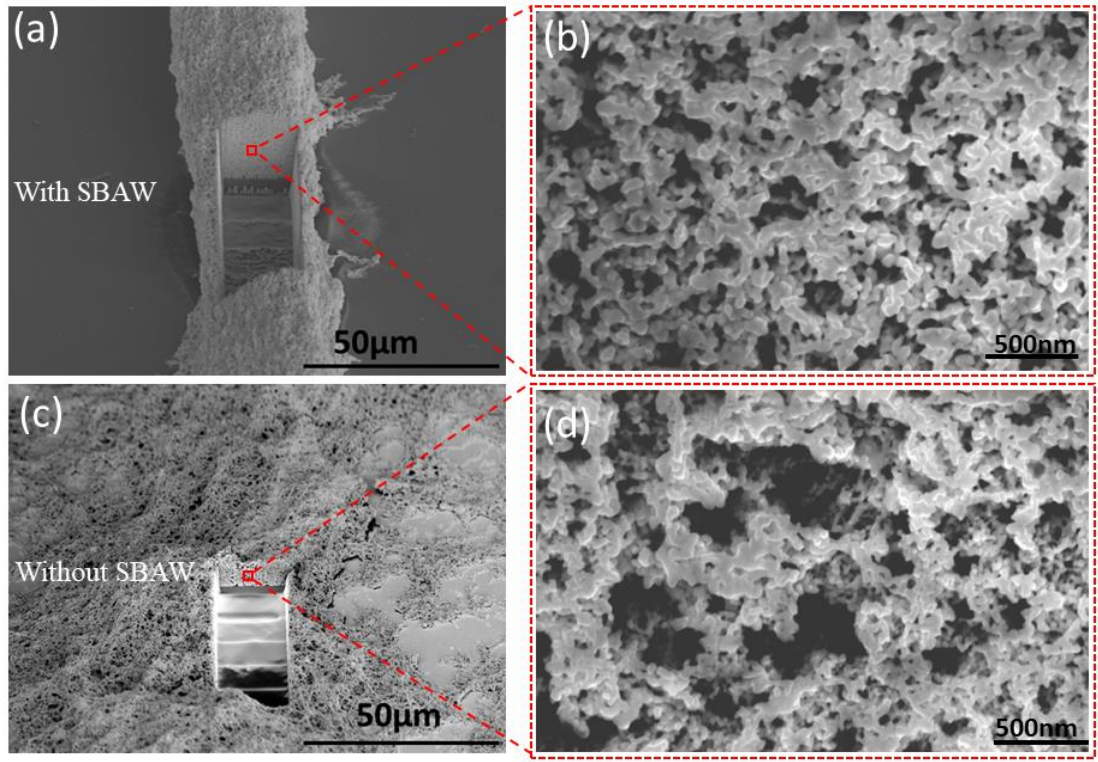

**Figure 5.** SEM images of (a) patterned gold structure and (c) reference sample. HR-SEM images of the cross sections (formed by FIB) of (b) patterned and (d) reference gold deposits.

Pore percentage calculations were performed by image analysis on five HR-SEM images. Although the densest areas of the reference samples were chosen, the pore percentage (31±4%) was ~50% higher for reference samples compared to rings formed by SBAWs (20±6%). This demonstrates that SBAWs affects both the mesostructure and the micro-structure. Gold and silver ring continuity are demonstrated by I-V measurements (**Figure S2**). The calculated average resistivity for gold was 2.5(±1.2)×10$^{-5}$ Ohm·m, three orders of magnitude higher than the bulk resistivity (2.20×10$^{-8}$ Ohm·m at room temperature),[43] as expected from the mesoporous structure.

Double layer and composite materials

After successfully demonstrating micro-fabrication of a single-layered material, the effect of standing acoustic waves was further examined on combinations of metals reduced from ions. Bimetallic structures were formed by reducing different metal ions both sequentially and simultaneously or by reducing metal ions in the presence of nano-additives.

To fabricate bimetallic layered structures (**Figure 6**a), 2 ml of silver nitrate solution were introduced into an acoustic resonator placed on single-sided polished silicon (~3×4 cm). Formation of silver NPs was observed after addition of sodium borohydride solution. The NPs were arranged almost immediately to form metallic ring shapes throughout the medium height following aggregation and sedimentation on the silicon substrate along wave nodes. After 30 min the aqueous medium was gently pumped out, leaving silver micro-patterns. 2 ml of gold chloride solution were introduced on top of the formed silver micro-structures. The solution was subjected to SBAWs followed by addition of borohydride solution. After 30 additional minutes to allow complete precipitation of formed particles the aqueous medium was evacuated by pipetting and samples were heated on a hot plate at 70°C for 10 min, washed with water and isopropanol and dried again (70°C for 1 min). As expected for particles with a positive acoustic factor, NPs were deposited at pressure nodes. To determine material distribution, cross-sectioning was performed. Single-sided polished silicon allows a more precise cut than glass.

The cleaving process is explained and shown in the supplementary information (**Figure S3**). Structures were analyzed by SEM and EDS (**Figure 7**b).

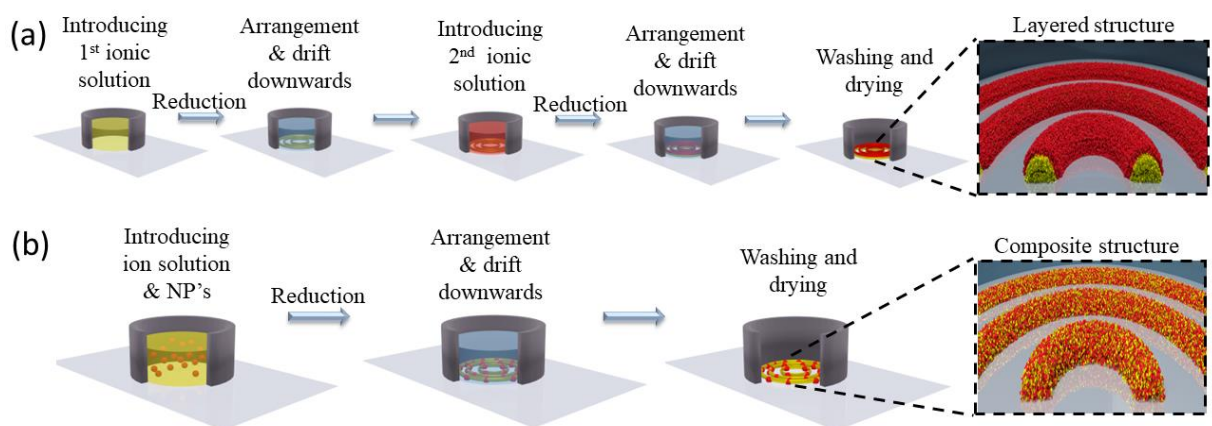

**Figure 6.** Illustration of micropattern fabrication steps for (a) multilayered products and (b) composite structures.

To fabricate a composite, a 2 ml solution of silver nitrate and gold chloride was introduced in the acoustic resonator placed on a single-sided polished silicon (~3×4 cm). After addition of sodium borohydride solution, formed NPs arranged almost immediately into metallic rings throughout the medium height followed by aggregation and sedimentation on the silicon substrate along wave nodes (Figure 6b). After 30 min, the aqueous medium was gently pumped out, and samples were heated on a hot plate at 70°C for 10 min, washed with water and isopropanol, and dried again (70°C for 1 min). Figure 7a shows a uniform composite structure obtained by simultaneous reduction of both metal ions with no substantial preference for specific location along the structure.

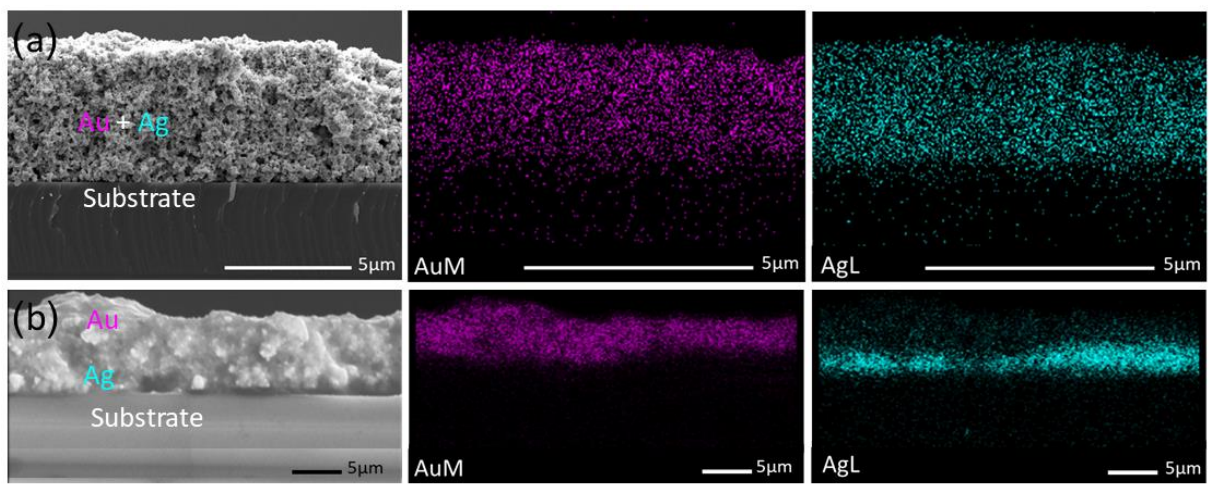

**Figure 7.** SEM and EDS images from cross-sections of formed micro-patterned structures. (a) Mixed Au/Ag composite when both ions are reduced simultaneously. (b) Au layer on top of Ag layer after consecutive reduction.

In addition to simultaneous synthesis and patterning, preformed NPs may be integrated into the structure formed by metal reduction. Composite metal/oxide-NP micro-structures were successfully fabricated by introducing hematite NPs into silver or gold reduction in the presence of acoustic waves (**Figure S4**). Bilayer structures were fabricated by introducing hematite NPs after formation of silver depositions (**Figure 8**a) and multi-layered patterns were achieved by deposition of reduced gold on top of the bilayer structure (Figure 8b).

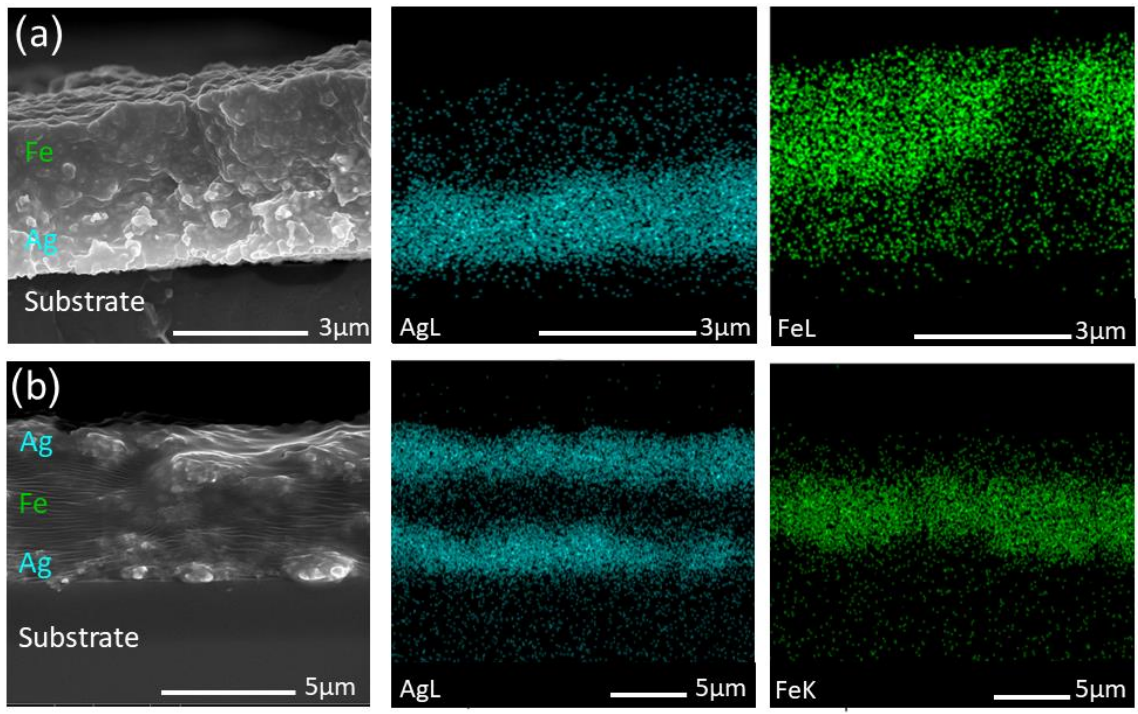

**Figure 8.** (a) HR-SEM and EDS images of cross-sectioned a bi-layered pattern of silver (bottom layer) and hematite NPs (top layer) and (b) multi-layered pattern of silver (top and bottom layers) and hematite NPs (middle layer).

Mixed composite structures were formed with controlled properties. A series of samples with varying hematite NP content were fabricated. I-V curves show negative correlation between sample conductivity and NP weight percentage (**Figure 9**) as expected from the electrical properties of hematite. Even for the highest concentration (12 wt%) mA currents were measured indicating that hematite NPs did not form large aggregates leading to electrical discontinuities. Careful consideration of the various combinations reveals that this straightforward approach holds significant promise for generating a multitude of composites.

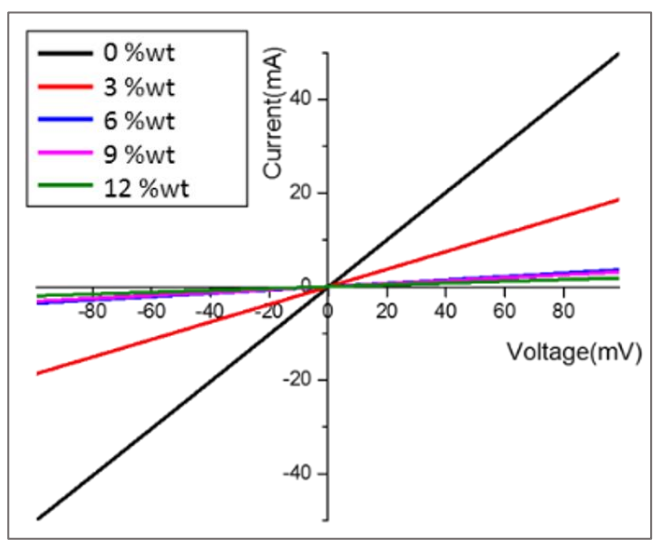

**Figure 9.** I-V curves of composite silver matrix with different hematite (Fe2O3) content.

CONCLUSIONS

In this study, SBAWs were employed to fabricate continuous conductive metal structures on a glass substrate by reducing metal ions in a cylindrical resonator. Metal ions were reduced in a cylindrical resonator, and the resultant NPs were simultaneously directed to specific locations, forming a metallic matrix with circular micro-patterns. The effect of the acoustic radiation force on the height of the deposit and the matrix density was demonstrated. This one-pot approach does not require stabilization of the formed NPs or specific treatment to enable agglomeration among them. Both metallic and multi-metallic layered or mixed composite micro-structures were demonstrated in fewer steps in comparison to other techniques. SBAWs enable us to work on large scales, paving the way for the fabrication of multifunctional sensors and catalysts.[44] Further research is required to identify additional potential applications that can benefit from this adaptable patterning technique.

ASSOCIATED CONTENT

**Supporting Information**.

The supporting Information is available.

Mathematical expression for acoustic radiation force in a cylindrical reservoir; Bright field microscope images of micro-patterned gold rings in a cylindrical reservoir; I-V curve of a gold micro-pattern; Setup and image of a pattern during cross-sectioning; Illustration of sample preparation and resultant SEM cross-section image of a bi-layered silver/gold ring on a silicon substrate; SEM and EDS images of $Fe_2O_3$ nanoparticles in a gold matrix composite; Bright field microscope images of concentric pattern fabricated through the reduction of various $HAuCl_4$ concentrations; Parallel silver stripes fabricated by a parallel transducers setup.

AUTHOR INFORMATION

**Corresponding Author**

Prof. Hagay Shpaisman, PhD, Department of Chemistry and Institute of Nanotechnology and Advanced Materials

Email: hagay.shpaisman@biu.ac.il

**Present Addresses**

Bar-Ilan University, Ramat Gan 5290002, Israel

**Author Contributions**

The manuscript was written through contributions of all authors. All authors have given approval to the final version of the manuscript.

**Funding Sources**

Israel Science Foundation grant No. 952/19

ACKNOWLEDGEMENTS

The authors acknowledge careful editing of the manuscript by Dr. Yuval Elias and financial support by the Israel Science Foundation grant No. 952/19.

TOC

A novel single-step fabrication method utilizes standing acoustic waves to guide the reduction products of metal ions, facilitating the creation of multi-layered and composite metallic micro-structures.

SYNOPSIS

For Table of Contents Only

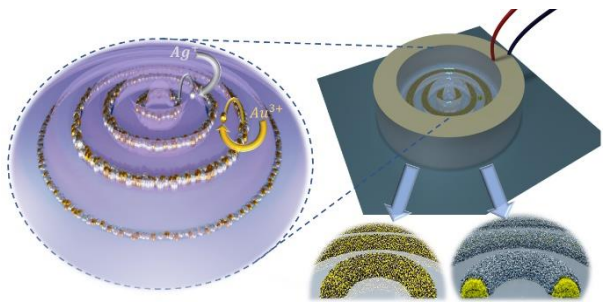

# Supporting Information

# One-Pot Approach for Acoustic Directed Assembly of Metallic and Composite Microstructures by Metal Ion Reduction

*Avraham Kenigsberg, Heli Peleg-Levy, Haim Sazan, Silvia Piperno, Liron Kenigsberg & Hagay Shpaisman\**

Department of Chemistry and Institute of Nanotechnology and Advanced Materials, Bar-Ilan University, Ramat Gan 5290002, Israel

E-mail: hagay.shpaisman@biu.ac.il

**S1 Micro-patterned gold rings.**
Figure S1 shows inverted bright field microscope images of typical acoustically micropatterned gold rings, from ring number two (from reservoir center) to sixteen (most exterior ring visible). The variation in the acoustic force, which acts as a Bessel function of the first kind (as described in Eq. S1 below) is reflected along the radii.

**Equation S1.**

$$F(r) = \frac{\pi R^3 \overline{p^2}}{3 \rho_m c_m^2} \left\{ \left[ 2\left(1 - \frac{\rho_m c_m^2}{\rho_p c_p^2}\right) + \frac{3(\rho_p - \rho_m)}{2\rho_p + \rho_m} \right] J_0(kr) - \frac{3(\rho_p - \rho_m)}{2\rho_p + \rho_m} J_2(kr) \right\} J_1(kr)$$

Where $J_0$, $J_1$ and $J_2$ are the $0^{th}$, $1^{st}$ and $2^{nd}$ order Bessel functions of the first kind, $\overline{p^2}$ is the time average pressure, $k$ is the wave number, $r$ is the radial coordinate originating at the center of the reservoir, and $\rho_m/\rho_p$ and $c_m/c_p$ are the medium/particle density and speed of sound. $R$ is the radius of a particle subject to the acoustic radiation force.

Rings with lower numbers are significantly denser and have more well-defined edges. Conversely, the number of non-patterned gold residues increases with ring number. This is in accord with weaker acoustic force close to the reservoir edge that cannot manipulate all reaction products; gravity and Brownian motion as well as Van-Der-Waals and drag forces are prominent in determining location of some of the deposits.

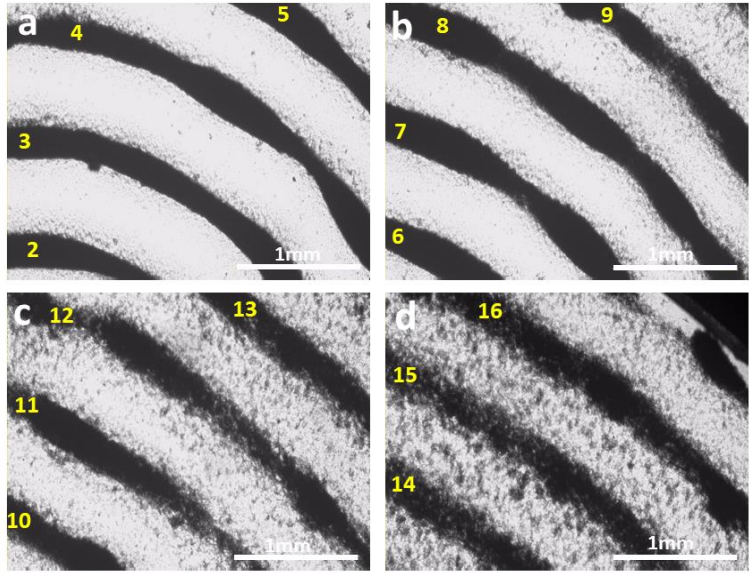

**Figure S10.** (a–d) Bright field microscope images of micro-patterned gold rings in a cylindrical reservoir. Ring numbers (starting from the reservoir center) are marked in yellow.

**S2 I-V measurements.**

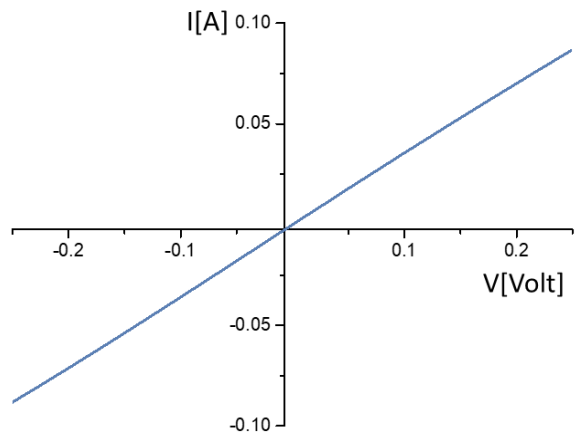

**Figure S2.** Typical I-V curve of a gold micro-pattern

**S3 Cross-section analysis.**
Micro-patterns were deposited on a polished silicon wafer, and cross-sections were performed using a LatticeAx 120 device (Figure S3 a, b). The wafer was fixed on a standing SEM stub at 90° such that the cross-section points upwards as illustrated in Figure S3 c. Figure S3 d shows a SEM image of a cross-section of a bi-metallic bi-layered silver/gold ring. The smooth silicon surface on the same plane with the micro-structure indicates a successful cross-sectioning (no/minimal structure distortion).

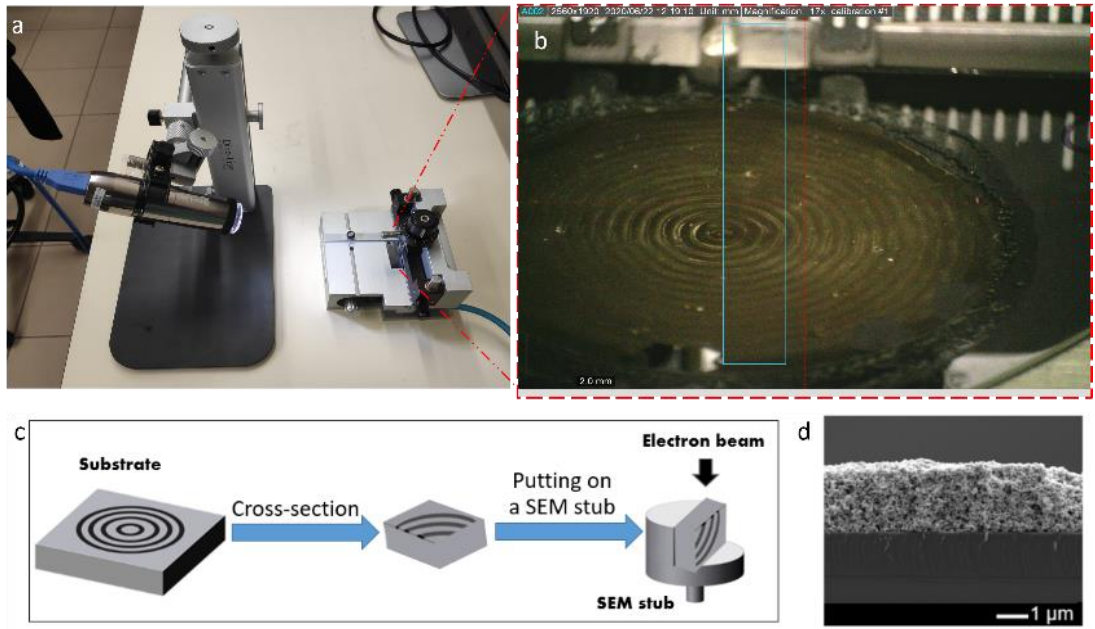

**Figure S3.** (a) Setup and (b) image of a pattern during cross-sectioning. (c) Illustration of sample preparation. (d) SEM cross-section image of a bi-layered silver/gold ring on a silicon substrate.

`S4 Composite structure of iron oxide NPs embedded in a gold matrix.`

Fe(III) oxide NPs were embedded in a concentric porous gold micro-pattern. This mixed composite structure was fabricated using acoustic waves by reduction of $HAuCl_4$ with $NaBH_4$ in alkaline solution while adding Fe(III) oxide NPs (<50 nm). Figure S4 shows SEM and EDS images of the acoustic-directed products.

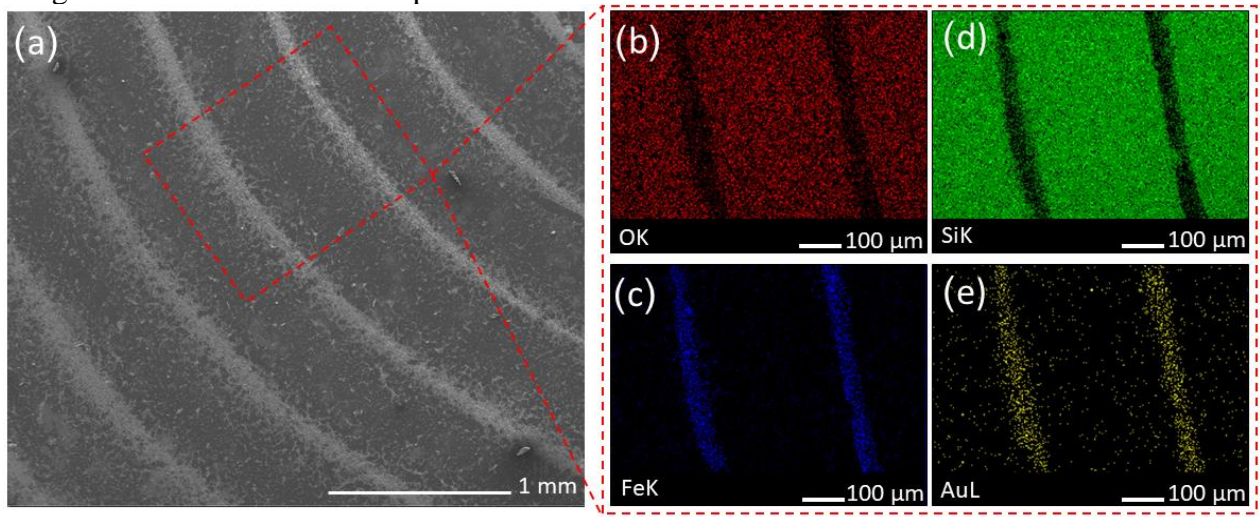

**Figure S4.** (a) SEM and (b–e) EDS images of $Fe_2O_3$ nanoparticles in a gold matrix composite.

`S5 Controlling the pattern width by tunning a limiting reagent.`

Figure S5 illustrates how altering the concentration of a limiting reagent ($HAuCl_4$) allows control of the pattern's dimensions. Using an inverted optical microscope, we measured the width of the second ring in each sample 30 minutes after adding the reducing agents. The average pattern's width increased from 50μm for 0.001M $HAuCl_4$ to 116μm and 160μm for concentrations of 0.002M and 0.003M, respectively.

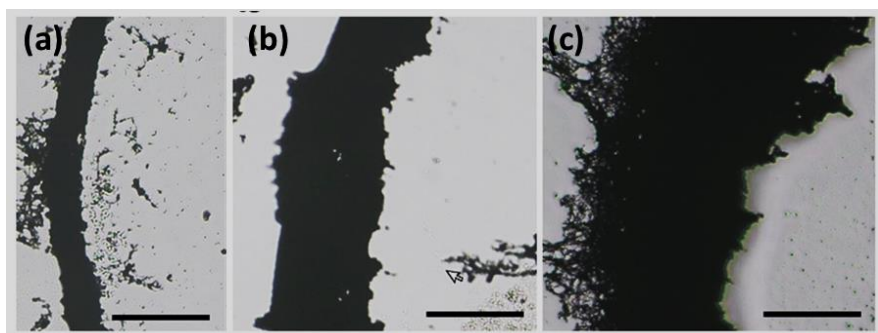

**Figure S5.** Bright field microscope images of the 2nd ring in a concentric pattern fabricated by reduction of (a) 0.001M, (b) 0.002M, (c) 0.003M $HAuCl_4$ in the presence of SBAWs in a cylindrical reservoir. Scale bar = 100μm

S6 Parallel silver micropatterns

Parallel silver micropatterns were created by generating standing waves within a square-shaped polycarbonate reservoir (23mm, 23mm, 11mm) using an acoustic patterning device equipped with two parallel piezo transducers (10, 30, 0.5mm, PZT, Americanpiezo). The aqueous solution volume was maintained at 2 ml, and the reagent concentration matched the method section's description for silver pattern fabrication.

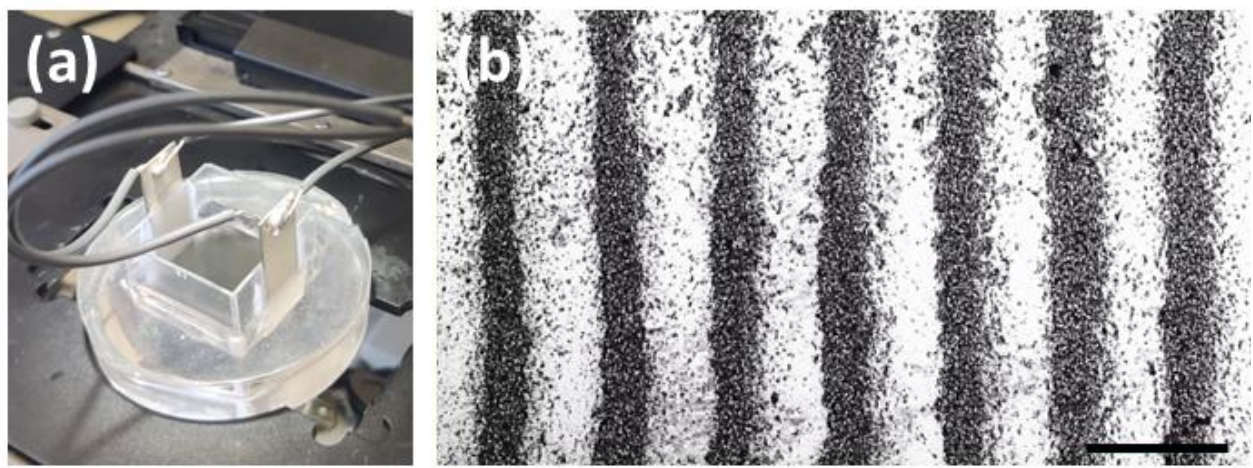

**Figure S6.** (a) Acoustic patterning device, with pair of parallel piezo transducers (resonance frequency: 4.6MHz) used to generate ultrasound standing waves across the square reservoir and (b) Parallel silver micropatterns. Scale bar = 200μm